\documentclass[aps, prb, floatfix, showpacs, twocolumn,amsmath,amssymb,]{revtex4-1}

\usepackage{graphicx}
\usepackage[latin1]{inputenc}
\usepackage{dcolumn}
\usepackage{bm}

\begin{document}

\title{Low-temperature transport properties of a two-dimensional electron gas with additional impurities}

\author{E. P. Rugeramigabo}
\email{rugeramigabo@nano.uni-hannover.de}
\affiliation{Institut f\"ur Festk\"orperphysik, Leibniz Universit\"at Hannover, Hannover, Germany}

\author{L. Bockhorn}
\affiliation{Institut f\"ur Festk\"orperphysik, Leibniz Universit\"at Hannover, Hannover, Germany}

\author{R. J. Haug}
\affiliation{Institut f\"ur Festk\"orperphysik, Leibniz Universit\"at Hannover, Hannover, Germany}

\date{\today}

\begin{abstract}

We investigate the effect due to background impurities embedded in the region of two-dimensional electron gases to the magnetotransport. These impurities are achieved by homogeneously incorporating Si atoms in single quantum wells of high quality GaAs/AlGaAs heterostructures. Several materials were grown this way with different densities of Si atoms. At low temperature the mobility decreases monotonically with increasing impurity density. In materials with incorporated impurities, effects are observed with increasing temperatures. On the one hand, a minimum in mobility appears, marking a transition between scattering enhancement due to acoustic phonon and mobility enhancement due to impurity driven conductivity. The temperature at which the transition takes place decreases with increasing impurity density. On the other hand, the incorporated impurities induce a non-monotonic temperature dependence of the electron density. The magnitude of this variation rises with increasing impurity density. The theory of electron-electron interaction correction to the conductivity alone fails to explain our results. 

\end{abstract}


\maketitle 

\section{Introduction}

The quality of two-dimensional electron gases (2DEG) located in GaAs/AlGaAs heterostructures is generally estimated by the zero-field electron mobility. In the range of mobilities higher than 10$^{6}\,$cm$^{2}$V$^{-1}$s$^{-1}$ this picture is not completely true, e.g regarding the regime of fractional quantum Hall effect. It was shown that the filling factor 5/2, expected for very high quality 2DEG, could be absent in very high mobility samples while occurring in slightly lower mobility ones \cite{Umansky2009}. Here, screening effects and interplay between different types of disorders play an important role \cite{Gamez2013}. Thus, the understanding of isolated scattering mechanisms in these high-mobility samples becomes an important issue.

The main scattering sources in GaAs/AlGaAs heterostructures are remote ionized impurities, interface roughness, alloy disorders, phonons and background impurities. Phonon scattering is a material specific property and cannot be addressed experimentally. Remote ionized donors are introduced by modulation doping \cite{Dingle1978} and provide electrons in the GaAs 2D-channel, what makes them unavoidable. Their scattering effect can however be considerably reduced by increasing the thickness of AlGaAs barrier or spacer, an action which results unfortunately in a reduction of the 2D-electron density. A way to overcome this restriction is to use a double heterostructure or quantum well (QW) structure. For the same spacer width one can double the electron density as compared to a simple heterostructure. Interface roughness and alloy disorders are negligible in high quality GaAs/AlGaAs and become an issue at high electron densities and very high mobilities. Background impurities are always present in epitaxial grown materials since they are inherent to the growth system and are spread all over the whole structure. The electrons interact with background impurities in the GaAs channel as well as with background impurities in the AlGaAs spacer. In this configuration it is not possible to separate the individual contributions. Most theoretical and experimental studies on the effect of background impurity scattering have been made at very low temperatures where this scattering mechanism is generally the limiting factor to the mobility. Extensive temperature dependent experiments taking also into account the exact location of the background impurities (in the electron channel or in the spacer) are still lacking.

\begin{figure} [t]
	\centering \includegraphics[clip,width=86mm]{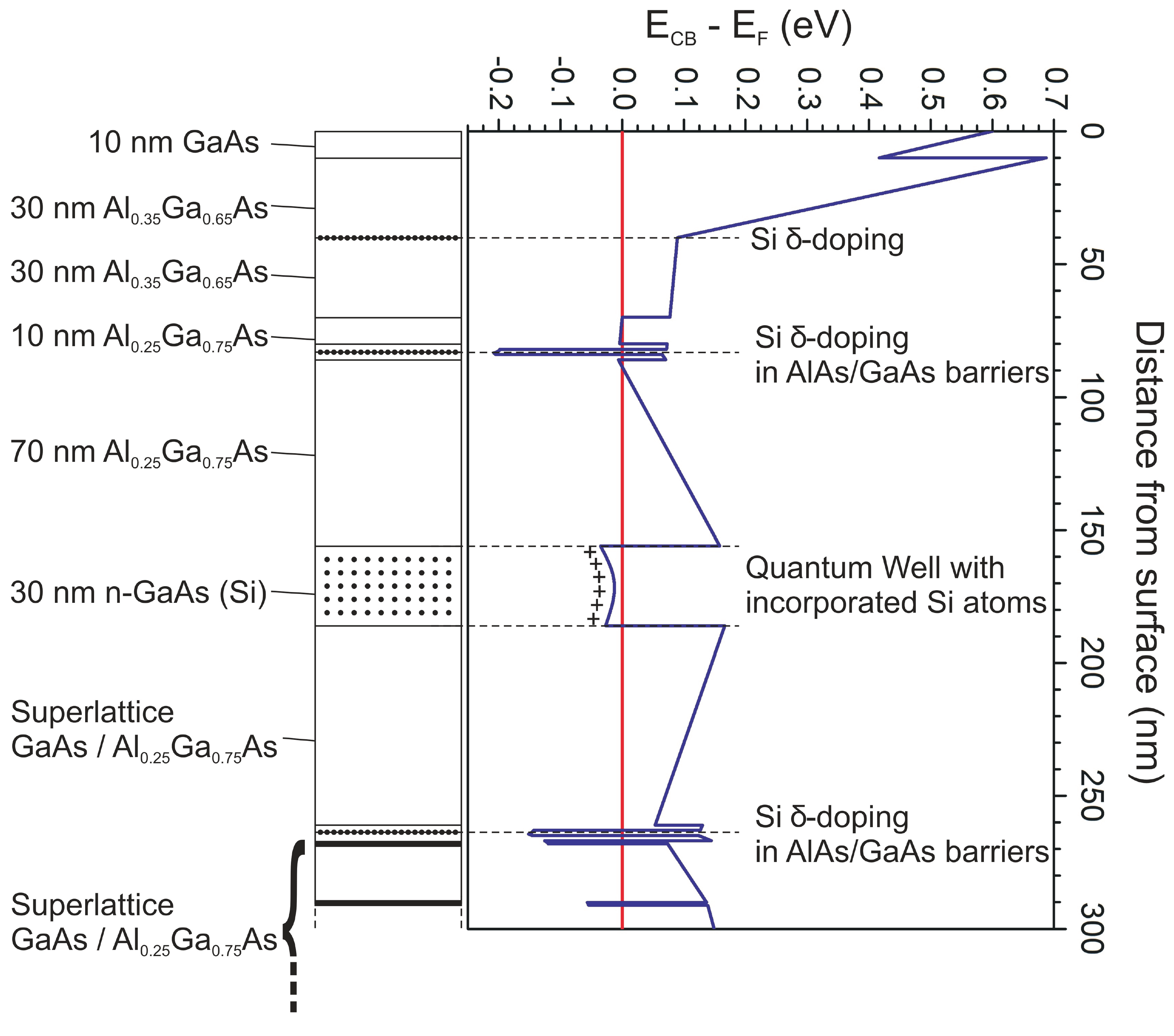}
	 \caption{\label{Band_structure} Calculated band diagram of the conduction band edge E$_{\texttt{CB}}$ relatively to the Fermi energy E$_{\texttt{F}}$ around the QW with incorporated Si impurities}
\end{figure}

In this study we report on the influence of the ionized background impurities in GaAs/AlGaAs QWs. In order to differentiate between ionized background impurities located within the 2DEG and those from the AlGaAs spacer region, impurities were intentionally incorporated only in the QW region. We investigate their effect on the temperature dependence of zero-field electron mobility $\mu_{E}$ and of the 2D~electron density $n_{E}$.

\section{Experimental setup \label{Sec:experimental} }

\begin{table}

\caption{\label{tab:parameters} Sample parameters determined at 50 mK: electron density $n_{E}$, low-field electron mobility $\mu_{E}$ and scattering rate $\tau_{imp}^{\rm -1}$ due to incorporated impurities of density $n_{imp}$ .}
\centering
\begin{ruledtabular}
\begin{tabular}{ccccc}
&\textbf{$n_{imp}$}&\textbf{$n_{E}$}&
\textbf{$\mu_{E}$}&\textbf{$\tau_{imp}^{\rm -1}$}\\

Sample&(10$^{15}$$\,$cm$^{-3}$)&(10$^{11}$$\,$cm$^{-2}$)&
(10$^{3}$$\,$cm$^{2}$V$^{-1}$s$^{-1}$)&(10$^{11}$$\,$s$^{-1}$)\\
\hline \\
A& 0 & 2.82	& 2020	& 0\\
B& 2.5 & 2.94 & 265 & 0.8\\
C& 5.1 & 2.95 & 128 & 1.8\\
D& 10 & 2.87 & 52 & 4.9\\
E& 40 & 3.66 & 11 & 19\\
\end{tabular}
\end{ruledtabular}
\end{table}

The GaAs/AlGaAs materials including a QW were grown by molecular beam epitaxy on GaAs (100) substrates. The relevant part of the material structure consists of a 30$\,$nm GaAs QW 150$\,$nm beneath the surface. 2D~electrons are provided in the QW by modulation doping from both sides due to Si $\delta$-doping layers with a sheet density $n_D$~=~1.6~$\cdot$~10$^{12}$cm$^{-2}$. The doping layers are located in 2$\,$nm thin GaAs layer sandwiched between 2$\,$nm AlAs barriers, forming a short period AlAs/GaAs superlattice. These are separated from the QW by a 70$\,$nm wide AlGaAs spacer layer. Further a Si $\delta$-doped layer has been inserted 40$\,$nm beneath the surface, in the middle of a 60$\,$nm AlGaAs layer, to screen the surface states. This structure has shown to be suitable for high-mobility GaAs/AlGaAs QW materials \cite{{Friedland1996},{Bockhorn2011}}. Additional Si-atoms were incorporated in the GaAs QW (described above), in which they act as impurities. This has been performed by co-evaporating Si and GaAs during the growth of the QW. Several materials were grown with the same layer structure but with different densities $n_{imp}$ of impurities in the QW. Fig.~\ref{Band_structure} shows the band structure for the relevant part around the QW for incorporated impurities \cite{{Snider1990},{Tan1990}}.

Samples are standard Hall bar geometries structured by photo-lithography and chemical wet etching. Low temperature magnetotransport measurements were carried out in a dilution refrigerator with a base temperature of 50$\,$mK. Additionally for temperature dependent measurements we used a $^4$He-cryostat equipped with a $^3$He-inset. All measurements were performed using a standard lock-in technique at 13$\,$Hz.

The samples and their corresponding $n_{imp}$ densities are given in Table \ref{tab:parameters}. Sample A is the reference sample without any incorporation of Si-atoms in the QW. The impurity density due to Si-atoms in the QW increases from sample B to sample E.

\section{Results and discussion \label{Sec:Results} }

\subsection{Low temperatures \label{low T}}

\begin{figure}[t]
   \centering
   \includegraphics[clip,width=80mm]{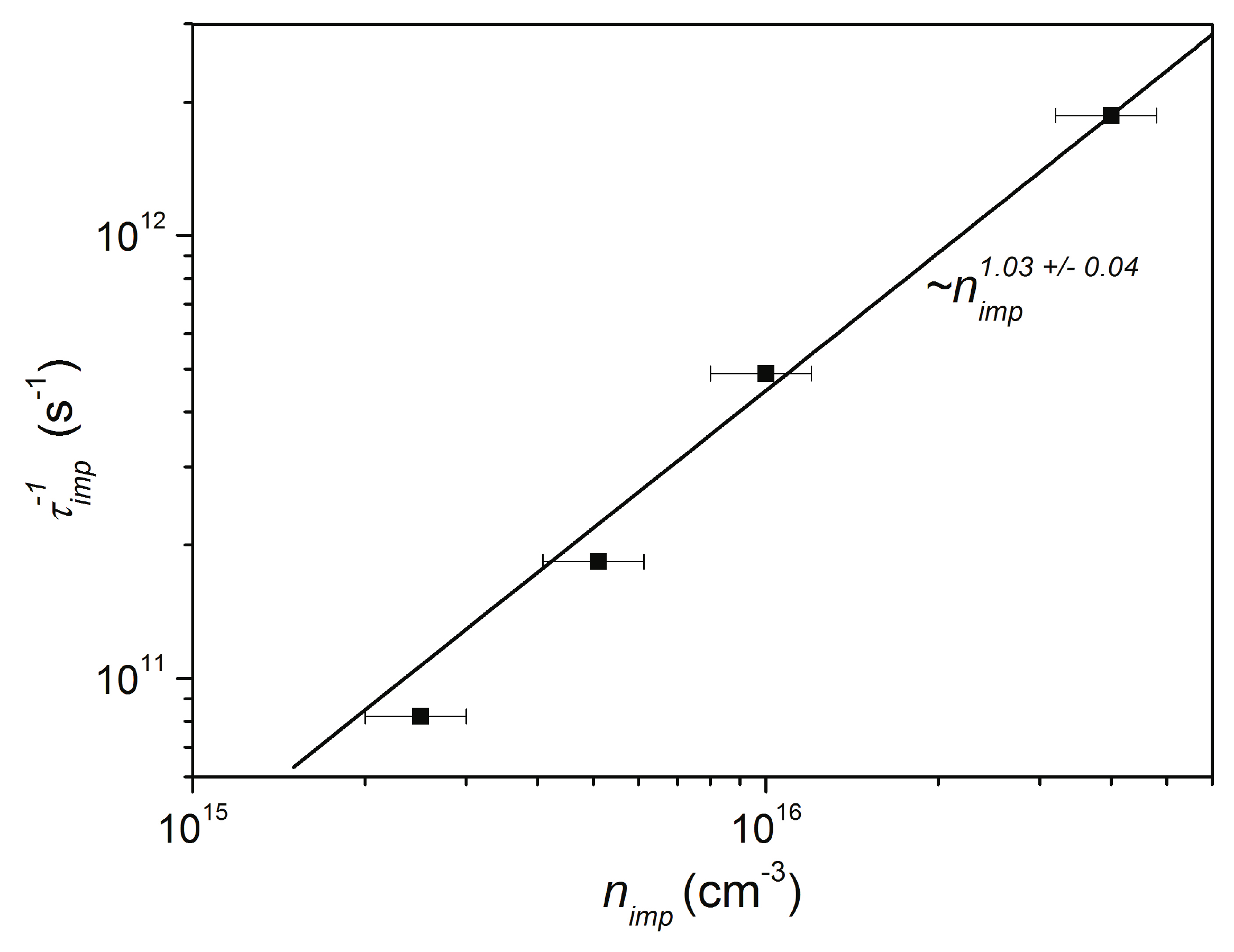}
	\centering
   \caption{\label{Tauimp} Double logarithmic plot of $\tau_{imp}^{\rm -1}$ is shown vs the density of the incorporated impurities $n_{imp}$. Squares with error bars represent the experimental data regarding Table \ref{tab:parameters}. The solid line is a fit of the form \mbox{$\tau_{imp}^{\rm -1}\,=\,\widetilde{A}_{fit}\,n_{imp}^{\beta_{fit}}$} with $\widetilde{A}$ and $B$ being constants}
\end{figure}

The characteristic values of all samples are given in Table \ref{tab:parameters} at 50$\,$mK . $n_{E}$ is calculated from $n_{E}$~=~1~/~($e R_{H}$),  where $R_{H}$ is the Hall coefficient obtained by fitting linearly the Hall resistance for 0.05~$T < B < B_{SdH}$. $B_{SdH}$ is the magnetic field at which the Shubnikov-de Hass ($SdH$) oscillations begin. Here it is worth to mention that for all samples ~0.05$\,$T$< B_{tr}$, the length scale of the weak localization magnetic field. From the $SdH$ oscillations we found the electron density $n_{SdH}$ equal to $n_{E}$. %As expected
All samples with incorporated impurities in the QW have higher electron densities than that of the reference sample, due probably to the doping effect of the incorporated Si impurities. 

Table \ref{tab:parameters} indicates that the reference sample has a good quality with a high mobility $\mu_{E}$ of about 2.02~$\cdot$~10$^{6}$$\,$cm$^{2}$V$^{-1}$s$^{-1}$. The mobility decreases with increasing impurity density. The samples were grown with an identical structure, so that the mobility drop can be ascribed to scattering events due to the rising number of impurities in the QW. In fact, using Matthiessen rule, we have 
\begin {equation}
	\tau_{tr}^{\rm -1} = \sum \tau_{i}^{\rm -1}. 	
		\label{eq:tau}
\end{equation}

$\tau_{tr}$ is the transport scattering time defined as $\tau_{tr}~=~m^{*} \mu_{E}/e$, with $m^{*}$ the effective electron mass and $e$ the elementary charge. $\tau_{i}$ represent scattering times from several types of disorders such as background impurities, remote ionized impurities, phonons, interface roughness and alloy disorders. In our samples, contributions from these different scattering processes to $\tau_{tr}$ is rather unknown and their determination is not the scope of this study. On account of this, care was taken to change only the density of impurities. Hence, equation~(\ref{eq:tau}) can be simplified to: 
\begin {equation}
	\tau_{tr}^{\rm -1} = \tau_{0}^{\rm -1} + \tau_{imp}^{\rm -1}~,
	\label{eq:tauimp}
\end{equation}

where $\tau_{imp}$ represent the scattering time due to the incorporated impurities in the GaAs QW and $\tau_{0}$ is the sum of all other scattering processes. The transport scattering time $\tau_{tr}^{\rm A}$ of the reference sample with $n_{imp}$=0 is equal to $\tau_{0}$. The calculated values of $\tau_{imp}^{\rm -1}$ are shown in Table \ref{tab:parameters}. The dependence of $\tau_{imp}^{\rm -1}$ on the incorporated impurity density $n_{imp}$ is shown in Fig.~\ref{Tauimp}. The scattering time $\tau_{imp}$ dominated by background impurities (defined as short-range scatterers) is generally assumed as \cite{Li2003} \mbox{$\tau_{imp}^{\rm -1}$\,=\,$h\,n_{imp}$\,/\,($m^{*} k_{F}^{\rm 3} a_{B}^{\rm 2})$}, where $h$ is Planck constant, $k_{F}$ is Fermi wave number and $a_{B}$ is the Bohr radius. This expression can be rewritten as \mbox{$\tau_{imp}^{\rm -1}\,=\,\widetilde{A}\,n_{imp}^{\beta}$} with \mbox{$\beta$\,=\,1} and \mbox{$\widetilde{A}\,=\,h\,/\,(m^{*} k_{F}^{\rm 3} a_{B}^{\rm 2})\,=\,4.51\,\cdot\,10^{-5}\,$cm$^{-3}$/s}, calculated using experimental data. In Fig.~\ref{Tauimp} a slope is fitted to the experimental data and we obtained \mbox{$\widetilde{A}_{fit}\,=\,1.37\,\cdot\,10^{-5}\,$cm$^{-3}$/s} and \mbox{$\beta_{fit}\,=\,1.03\,\pm\,0.04$}. The observed proportionality between $\tau_{imp}^{\rm -1}$ and $n_{imp}$ is consistent with the expression above. We find that $\widetilde{A}_{fit}$ just differs from $\widetilde{A}$ by a numerical factor $A^{*}$ defined by \mbox{$\widetilde{A}_{fit}$\,=\,$\widetilde{A}\,A^{*}$} with \mbox{$A^{*}\,\approx\,0.3$}. The theoretical introduced factor by Dmitriev \textit{et al.} \cite{Dmitriev2012} (\mbox{$ln(k_{F}\,a_{B})$\,=\,0.296}) is in very good agreement with our determinant $A^{*}$. Therefore this reinforces our assumption that in the limit of our experimental errors the drop of the electron mobility with impurity density is directly linked to the incorporated impurities in the GaAs QW.

\subsection{Temperature dependence \label{high T}}

In order to understand the transport properties of our samples at higher temperatures, we performed temperature dependent magnetotransport measurements. With increasing temperatures the background impurity scattering is not anymore the dominant scattering mechanism. The increasing contribution from phonon (acoustic and polar optical) scattering with temperature as well as temperature dependent screening effects lead to interesting behavior in the variation of the mobility and electron density with temperature.

\subsubsection{Mobility \label{subseq:Mob}}

For high mobility sample at low temperatures (in the so-called Bloch-Gr\"uneisen regime \cite{Stormer1990}), the mobility is weakly temperature dependent and is mainly limited by the contribution $\mu_{d}$ due to scattering on ionized impurities and interface roughness. In fact a contribution to the mobility depending on acoustic phonons $\mu_{ac}$ exists, but in this regime $\mu_{d} << \mu_{ac}$ is valid. While $\mu_{d}^{\rm -1}$ is temperature independent, $\mu_{ac}^{\rm -1}$(T) has a temperature $T^{-5}$-dependence for piezoelectric interaction and a stronger temperature dependence $T^{-7}$ in the case of deformation potential. For temperature higher than the Bloch-Gr\"uneisen temperature regime $\mu_{ac}^{\rm -1}$(T) becomes linear and smaller compared to $\mu_{d}^{\rm -1}$. Up to 40$\,$K the importance of acoustic phonon scattering is predominant and will be overtaken by electron scattering on polar optical phonons at $T >$~40$\,$K. 

\begin{figure}[t]
	\centering
	\includegraphics[clip,width=80mm]{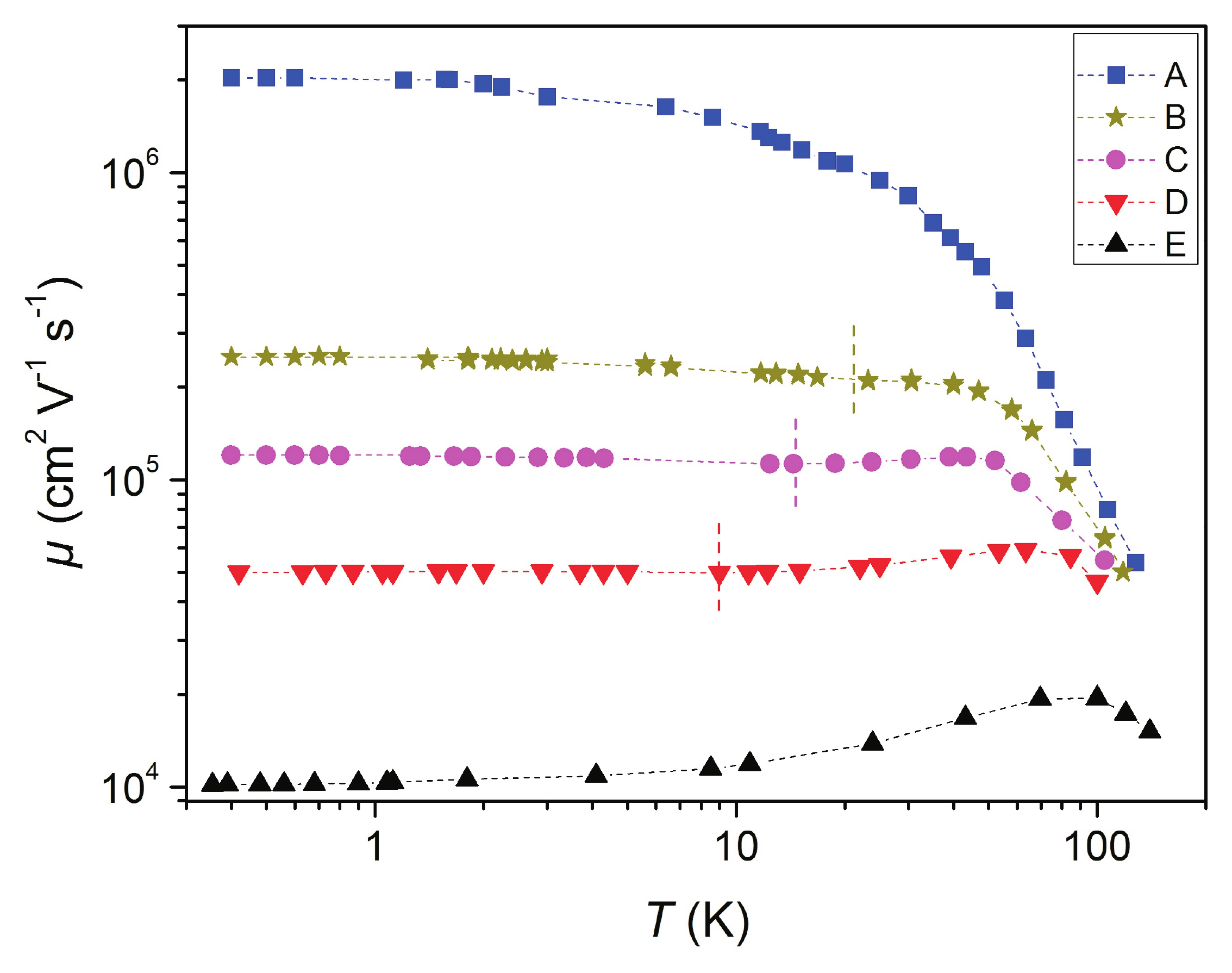}
	\centering
	\caption{\label{Mobility} Temperature dependence of the zero magnetic field mobility for samples A, B, C, D and E. Dashed lines are guides to the eye. The short vertical traces mark minima in the temperature dependence of the mobility.}
\end{figure} 

Fig.~\ref{Mobility} shows the temperature dependence of the mobility $\mu_{E}$(T) for the considered samples. The reference sample (A) shows a typical behavior for high mobility samples. In the range 4$\,$K$\,<T<\,$40$\,$K the mobility decreases due to scattering on acoustic phonons. This is corroborated by the linear dependence of the inverse mobility to the temperature $\mu^{-1}$~=~$\mu_{0}^{\rm -1}$~+~$\alpha_{ref}\,T$. We determine from data $\alpha_{ref}$~=~2.4~x~10$^{-8}\,$V~s~cm$^{-2}$~K$^{-1}$ which is usual for acoustic phonon scattering \cite{{Mendez1984},{Hirakawa1986},{Kawamura1992}}. The stronger decrease of the mobility for $T >$~40$\,$K is related to polar optical phonon scattering. This mobility drop at high temperatures is common for all considered samples. However, substantial differences appear in the behavior of $\mu_{E}$(T) for $T <$~40$\,$K when comparing the reference sample to the samples with impurities. The strongest difference is found for the highest impurity density sample E. Here the mobility increases with temperature in two steps: First, there is a weak mobility increase up to $T \approx$~4$\,$K, which is followed by a stronger one. This opposite temperature behavior of samples A and E are linked together, if one considers the behavior in samples with intermediate impurity densities. For sample B, C and D $\mu_{E}$(T) presents a non-monotonic behaviour. First it decreases similar to sample A with increasing temperature. At a temperature $T_{m}$, highlighted by short dashed vertical lines in Fig.~\ref{Mobility}, the mobility reaches a minimum and increases thereafter with distinct slopes for samples B, C and D. 

We see a clear trend if we consider $\mu_{E}$(T) at $T\,>\,T_{m}$. On the one hand, the lower the density of the incorporated impurity in the QWs the higher the temperature $T_{m}$ at which the mobility minimum is seen. On the other hand, the slope of the increasing mobility with temperature is found to be enhanced with the impurity density. Both effects point towards the fact that the temperature dependence of the mobility is clearly dominated by a temperature enhanced impurity mobility $\mu_{imp}$ for $T_{m}\,<\,T\,<$~40K. The situation is different at $T\,<T_{m}$ where the decrease of $\mu_{E}$ with increasing temperature infers a dominant contribution of acoustic-phonon scattering to the mobility. In fact the inverse mobility shows clearly a linear temperature dependence $\mu^{-1}$~=~$\mu_{0}^{\rm -1}$~+~$\alpha_{imp}\,T$. Intriguingly, we obtained an identical factor $\alpha_{imp}$~=~4.5~x~10$^{-8}$V~s~cm$^{-2}$~K$^{-1}$ for the samples B, C and D. The value of the slope is higher than that of the reference sample. One reason might be that the contribution $\mu_{imp}$ of the ionized background scattering to the mobility has already a positive temperature dependence at low temperatures. As a consequence of this, a variation of the impurity density would result in a change of the behavior of $\mu_{imp}\,(T)$. This would lead to different values of $\alpha_{imp}$ for samples B, C and D, in contradiction to our findings. Here, it seems more reasonable to assume a temperature dependence of $\mu_{imp}$ first above $T_{m}$, so it is not responsible for the increase in $\alpha_{imp}$. As the samples have similar electron densities, the change from $\alpha_{ref}$ to $\alpha_{imp}$ can be related to a change in the deformation potential constant \cite{{Mendez1984},{Hirakawa1986},{Kawamura1992}} (which can vary between 7 and 16~eV) and/or a change in the spatial extension of the electronic wave function \cite{Stern1972} in the GaAs QW. A detailed theoretical calculation is needed to understand this increase as a direct consequence of the incorporation of impurity in the GaAs QW.  

The contribution of the acoustic phonon scattering to the mobility becomes first negligible when impurities in the QW reach a density $n_{imp} >$~1~$\cdot$~10$^{16}$~cm$^{-3}$. This can be related to an effective screening of the long range character of ionized impurities within the QW. Further increase in impurity density leads to a reduction of this screening effect. This is clearly seen in the mobility dependence of sample E (Fig.~\ref{Mobility}) where the mobility $\mu_{E}$ is already dominated by ionized impurity scattering, even at low temperatures.

\subsubsection{Electron density \label{subseq:2DEG}}

\begin{figure}[t]
   \centering
   \includegraphics[clip,width=86mm]{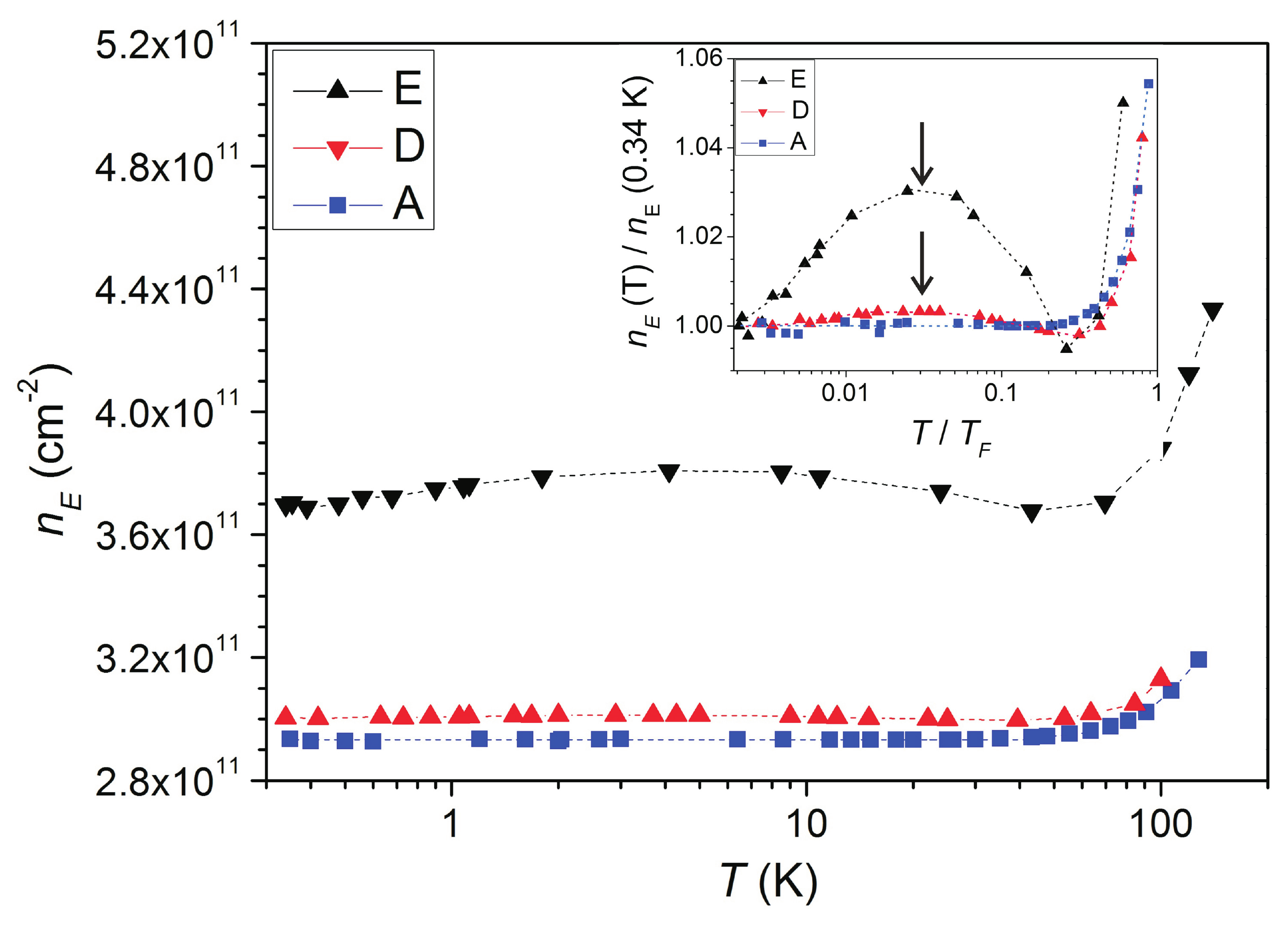}
	\centering
   \caption{\label{electron_density} Variation of the electron density with temperature for sample A, sample D and sample E. The inset shows electron density scaled over the density at $T =$~0.34$\,$K as a function of $T/T_{F}$. The arrows show the density maxima.}
\end{figure}

The determination of the electron density $n_{E}$ from the Hall coefficient $R_{H}$ performed at low temperature was also extended at higher temperatures. Fig.~\ref{electron_density} shows differences in the density $n_{E}$ with temperature for the samples A, D and E. We observe in all samples a strong increase of the electron density with temperature for $T >$~50$\,$K. This is supposed to be related to temperature induced ionization effects. At low temperatures ($T <$~50$\,$K), the reference sample A has a temperature independent electron density as observed quite commonly in high-mobility samples. For samples D and E however, we observe a non-monotonic temperature dependence of the electron density. With increasing temperature the electron density first increases (defined as \textit{region~I}) and then decreases (defined as \textit{region~II}). The slopes for both increase and decrease of the electron density with temperature become sharper with increasing impurity density. Samples B and C (not shown) don't show such a temperature dependence of the electron density.

The incorporated impurities form impurity bands (IB) in the band gap. In the case of a GaAs QW, Serre et al. \cite{Serre1989} have shown (using a model of a delta-doping in the QW which can be generalized to a homogeneous doping as in our case) that the binding energy is not constant. The binding energy is maximum in the middle and minimum at the edges of the QW. At a certain impurity density, the impurity band merges into the conduction band (CB). In the case $n_{E} >> n_{imp}$ the merging of the IB into CB can occur even at lower impurity densities. So, the problem may be reduced to the occupation probability of the Fermi level if, due to the broadening of the impurity band, states are available continuously across the Fermi level. 

In the inset of Fig.~\ref{electron_density} the dependence on $T/T_{F}$ is shown for the considered $n_{E}$ in Fig.~\ref{electron_density} which have been additionally normalized to the value at the lowest temperature of 0.34$\,$K for more clarity. $T_{F}$ is the Fermi temperature and was determined by using the maximum density (between \textit{region~I} and \textit{region~II}). Clearly, the maximum is found at the same $T/T_{F}$ value of 0.03. Interestingly, the only remaining difference in the temperature dependence lies only in the magnitude of the density variation with temperature. Thus, the same process obviously leads in both samples to a variation of the electron density with temperature. %So, the difference in magnitude can be directly related to the density of incorporated impurity in the QW.

The temperature dependence of $R_{H}$ (and thus of $n_{E}$) is directly related to the temperature and magnetic field dependence of the conductivity tensor. Due to the presence of impurities, corrections to the Drude conductivity are introduced. At low magnetic fields the main contributions arise from the electron-electron interaction ($\textit{EEI}$) and from weak localization ($\textit{WL}$) \cite{{Altshuler1985},{Lee1985}}. As mentioned before the weak localization contribution to the conductivity was neglected by using an adequate magnetic field range. The $\textit{EEI}$ correction consists of a diffusion regime for $k_{B}\,T\,\tau/\hbar<<1$ and a ballistic one for $k_{B}\,T\,\tau/\hbar>>1$. But in both regimes and in our range of electron densities, the decrease of the electron density with increasing temperature is not predicted by the $\textit{EEI}$ theory. So the $\textit{EEI}$ analysis will be restricted to \textit{region~I} and it is suitable to focus first on the diffusion regime. In fact the $\textit{EEI}$ correction in this regime affects only the longitudinal conductivity $\sigma_{xx}$, leaving the diagonal conductivity $\sigma_{xy}$ unchanged \cite{{Zala2001},{Zala2001a}}. 
\begin{equation}
		\sigma_{xx}(T,B) = \frac{e\,n\,\mu_{D}(T)}{1+\mu_{D}^{2}(T)\,B^{2}}+\delta \sigma_{xx}^{\rm ee}(T),
	\label{eq:1}
\end{equation}
\begin{equation}
	\sigma_{xy}(T,B) = \frac{e\,n\,\mu_{D}^{2}(T)\,B}{1+\mu_{D}^{2}(T)\,B^{2}}
	\label{eq:2}
\end{equation} 
where $\mu_{D} $ is the Drude mobility. $\sigma_{xx}(T,B)$ = $a/(a^{2}$+$b^{2}$) and $\sigma_{xy}(T,B)$= $b/(a^{2}$+$b^{2}$) with $a = \rho_{xx}(T,B)$ being the longitudinal resistivity and $b = \rho_{xy}(T,B)$ being the diagonal resistivity. $\delta \sigma_{xx}^{\rm ee}$ is the diffusive part of the $\textit{EEI}$ correction to the conductivity.

\begin{figure}[t]
   \centering
   \includegraphics[clip,width=80mm]{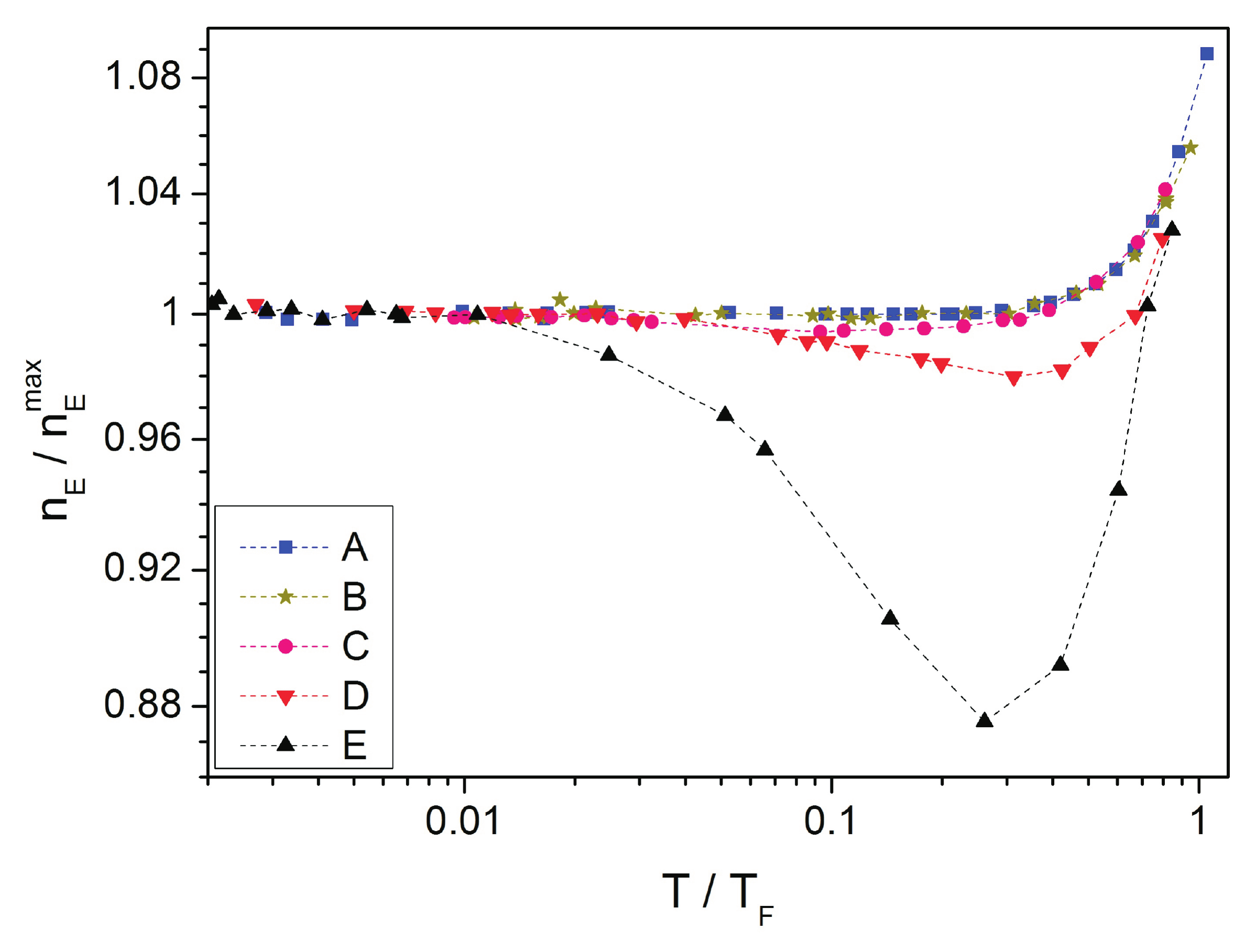}
	\centering
   \caption{\label{electron_density_normed}$T/T_{F}$ dependence of the normalized 2D electron density over the maximum density for samples A, B, C, D and E}
\end{figure}

This analysis can be performed on samples D and E which satisfy the condition $k_{B}\,T\,\tau/\hbar\,<<\,1$ for $T/T_{F}\,<<\,0.03$ and $T/T_{F}\,<<\,0.1$, respectively. We fitted the experimental data with $n$ and $\mu_{D}(T)$ as fit parameters. In the considered temperature range (\textit{region~I}) we obtain a constant 2D~electron density $n_{E}$ of 3.04~$\cdot$~10$^{11}$$\,$cm$^{-2}$ for sample D and 4~$\cdot$~10$^{11}$$\,$cm$^{-2}$ for sample E. The corrected values for samples D and E together with the electron densities of the other considered samples are shown in Fig.~\ref{electron_density_normed} normalized to the maximum electron density at low temperature as function of $T/T_{F}$. Obviously $n_{E}$ is constant in all samples up to $T/T_{F}\,\approx\,0.01$ and begins afterwards to decrease with a slope that increases with impurity density in \textit{region~II}. 

With the assumption of an energy dependent mobility, we performed thermal averaging of the energy around the Fermi level. While this calculation reflects also a decrease of the electron density with temperature, the slope is stronger than the measured temperature dependence. A reason for this discrepancy might be that the changing screening due to the QW impurity density and to temperature is not correctly taken into account. 
Since the temperature dependence of the electron density in \textit{region~I} and \textit{II} takes place at $T << T_{F}$ we need a unifying theory. Ionized background impurity screening might play a significant role, because the electron density variation is directly related to the Fermi temperature. An extension of the screening theory developed in Ref. \cite{DasSarma2005} for low electron density systems to high density disordered systems may find application to our system.

\section{Conclusion}

The effect of background impurities located in the 2DEG has been investigated. The used materials were realized by intentionally incorporating impurities in form of Si atoms in the QW of high-mobility modulation-doped GaAs/AlGaAs heterostructures. For all the samples we showed that at low temperatures the mobility is limited by background impurity scattering. At intermediate temperatures a clear signature of the effect of incorporated impurities was seen. Upon incorporation and subsequent increase of impurities in the GaAs QW we observe a transition from acoustic phonon scattering (reference sample A) to background impurity enhanced conductivity (sample E) as the dominant contribution to the mobility. The impurities induced also a non-monotonic temperature dependence of the electron density. The low temperature increase in electron density with temperature can be explained by the theory of electron-electron interaction correction to the conductivity. The following decrease of the electron density with temperature was only qualitatively described by thermal averaging of the energy.

\begin{acknowledgements}
We are grateful to I. Gornyi and D. Polyakov for useful discussions. This work was financially supported by the Cluster of Excellence QUEST. A technical support from the Laboratory for Nano and Quantum Engineering is also acknowledged.
\end{acknowledgements}


\begin{thebibliography}{20}%
	\makeatletter
	\providecommand \@ifxundefined [1]{%
		\@ifx{#1\undefined}
	}%
	\providecommand \@ifnum [1]{%
		\ifnum #1\expandafter \@firstoftwo
		\else \expandafter \@secondoftwo
		\fi
	}%
	\providecommand \@ifx [1]{%
		\ifx #1\expandafter \@firstoftwo
		\else \expandafter \@secondoftwo
		\fi
	}%
	\providecommand \natexlab [1]{#1}%
	\providecommand \enquote  [1]{``#1''}%
	\providecommand \bibnamefont  [1]{#1}%
	\providecommand \bibfnamefont [1]{#1}%
	\providecommand \citenamefont [1]{#1}%
	\providecommand \href@noop [0]{\@secondoftwo}%
	\providecommand \href [0]{\begingroup \@sanitize@url \@href}%
	\providecommand \@href[1]{\@@startlink{#1}\@@href}%
	\providecommand \@@href[1]{\endgroup#1\@@endlink}%
	\providecommand \@sanitize@url [0]{\catcode `\\12\catcode `\$12\catcode
		`\&12\catcode `\#12\catcode `\^12\catcode `\_12\catcode `\%12\relax}%
	\providecommand \@@startlink[1]{}%
	\providecommand \@@endlink[0]{}%
	\providecommand \url  [0]{\begingroup\@sanitize@url \@url }%
	\providecommand \@url [1]{\endgroup\@href {#1}{\urlprefix }}%
	\providecommand \urlprefix  [0]{URL }%
	\providecommand \Eprint [0]{\href }%
	\providecommand \doibase [0]{http://dx.doi.org/}%
	\providecommand \selectlanguage [0]{\@gobble}%
	\providecommand \bibinfo  [0]{\@secondoftwo}%
	\providecommand \bibfield  [0]{\@secondoftwo}%
	\providecommand \translation [1]{[#1]}%
	\providecommand \BibitemOpen [0]{}%
	\providecommand \bibitemStop [0]{}%
	\providecommand \bibitemNoStop [0]{.\EOS\space}%
	\providecommand \EOS [0]{\spacefactor3000\relax}%
	\providecommand \BibitemShut  [1]{\csname bibitem#1\endcsname}%
	\let\auto@bib@innerbib\@empty
	%</preamble>
	\bibitem [{\citenamefont {Umansky}\ \emph {et~al.}(2009)\citenamefont
		{Umansky}, \citenamefont {Heiblum}, \citenamefont {Levinson}, \citenamefont
		{Smet}, \citenamefont {N\"ubler},\ and\ \citenamefont {Dolev}}]{Umansky2009}%
	\BibitemOpen
	\bibfield  {author} {\bibinfo {author} {\bibfnamefont {V.}~\bibnamefont
			{Umansky}}, \bibinfo {author} {\bibfnamefont {M.}~\bibnamefont {Heiblum}},
		\bibinfo {author} {\bibfnamefont {Y.}~\bibnamefont {Levinson}}, \bibinfo
		{author} {\bibfnamefont {J.}~\bibnamefont {Smet}}, \bibinfo {author}
		{\bibfnamefont {J.}~\bibnamefont {N\"ubler}}, \ and\ \bibinfo {author}
		{\bibfnamefont {M.}~\bibnamefont {Dolev}},\ }\href {\doibase
		http://dx.doi.org/10.1016/j.jcrysgro.2008.09.151} {\bibfield  {journal}
		{\bibinfo  {journal} {Journal of Crystal Growth}\ }\textbf {\bibinfo {volume}
			{311}},\ \bibinfo {pages} {1658 } (\bibinfo {year} {2009})}\BibitemShut
	{NoStop}%
	\bibitem [{\citenamefont {Gamez}\ and\ \citenamefont
		{Muraki}(2013)}]{Gamez2013}%
	\BibitemOpen
	\bibfield  {author} {\bibinfo {author} {\bibfnamefont {G.}~\bibnamefont
			{Gamez}}\ and\ \bibinfo {author} {\bibfnamefont {K.}~\bibnamefont {Muraki}},\
	}\href {\doibase 10.1103/PhysRevB.88.075308} {\bibfield  {journal} {\bibinfo
		{journal} {Phys. Rev. B}\ }\textbf {\bibinfo {volume} {88}},\ \bibinfo
	{pages} {075308} (\bibinfo {year} {2013})}\BibitemShut {NoStop}%
\bibitem [{\citenamefont {Dingle}\ \emph {et~al.}(1978)\citenamefont {Dingle},
	\citenamefont {St\"ormer}, \citenamefont {Gossard},\ and\ \citenamefont
	{Wiegmann}}]{Dingle1978}%
\BibitemOpen
\bibfield  {author} {\bibinfo {author} {\bibfnamefont {R.}~\bibnamefont
		{Dingle}}, \bibinfo {author} {\bibfnamefont {H.~L.}\ \bibnamefont
		{St\"ormer}}, \bibinfo {author} {\bibfnamefont {A.~C.}\ \bibnamefont
		{Gossard}}, \ and\ \bibinfo {author} {\bibfnamefont {W.}~\bibnamefont
		{Wiegmann}},\ }\href {\doibase http://dx.doi.org/10.1063/1.90457} {\bibfield
	{journal} {\bibinfo  {journal} {Applied Physics Letters}\ }\textbf {\bibinfo
		{volume} {33}},\ \bibinfo {pages} {665} (\bibinfo {year} {1978})}\BibitemShut
{NoStop}%
\bibitem [{\citenamefont {Friedland}\ \emph {et~al.}(1996)\citenamefont
	{Friedland}, \citenamefont {Hey}, \citenamefont {Kostial}, \citenamefont
	{Klann},\ and\ \citenamefont {Ploog}}]{Friedland1996}%
\BibitemOpen
\bibfield  {author} {\bibinfo {author} {\bibfnamefont {K.-J.}\ \bibnamefont
		{Friedland}}, \bibinfo {author} {\bibfnamefont {R.}~\bibnamefont {Hey}},
	\bibinfo {author} {\bibfnamefont {H.}~\bibnamefont {Kostial}}, \bibinfo
	{author} {\bibfnamefont {R.}~\bibnamefont {Klann}}, \ and\ \bibinfo {author}
	{\bibfnamefont {K.}~\bibnamefont {Ploog}},\ }\href {\doibase
	10.1103/PhysRevLett.77.4616} {\bibfield  {journal} {\bibinfo  {journal}
		{Phys. Rev. Lett.}\ }\textbf {\bibinfo {volume} {77}},\ \bibinfo {pages}
	{4616} (\bibinfo {year} {1996})}\BibitemShut {NoStop}%
\bibitem [{\citenamefont {Bockhorn}\ \emph {et~al.}(2011)\citenamefont
	{Bockhorn}, \citenamefont {Barthold}, \citenamefont {Schuh}, \citenamefont
	{Wegscheider},\ and\ \citenamefont {Haug}}]{Bockhorn2011}%
\BibitemOpen
\bibfield  {author} {\bibinfo {author} {\bibfnamefont {L.}~\bibnamefont
		{Bockhorn}}, \bibinfo {author} {\bibfnamefont {P.}~\bibnamefont {Barthold}},
	\bibinfo {author} {\bibfnamefont {D.}~\bibnamefont {Schuh}}, \bibinfo
	{author} {\bibfnamefont {W.}~\bibnamefont {Wegscheider}}, \ and\ \bibinfo
	{author} {\bibfnamefont {R.~J.}\ \bibnamefont {Haug}},\ }\href {\doibase
	10.1103/PhysRevB.83.113301} {\bibfield  {journal} {\bibinfo  {journal} {Phys.
			Rev. B}\ }\textbf {\bibinfo {volume} {83}},\ \bibinfo {pages} {113301}
	(\bibinfo {year} {2011})}\BibitemShut {NoStop}%
\bibitem [{\citenamefont {Snider}\ \emph {et~al.}(1990)\citenamefont {Snider},
	\citenamefont {Tan},\ and\ \citenamefont {Hu}}]{Snider1990}%
\BibitemOpen
\bibfield  {author} {\bibinfo {author} {\bibfnamefont {G.~L.}\ \bibnamefont
		{Snider}}, \bibinfo {author} {\bibfnamefont {I.-H.}\ \bibnamefont {Tan}}, \
	and\ \bibinfo {author} {\bibfnamefont {E.~L.}\ \bibnamefont {Hu}},\ }\href
{\doibase http://dx.doi.org/10.1063/1.346443} {\bibfield  {journal} {\bibinfo
		{journal} {Journal of Applied Physics}\ }\textbf {\bibinfo {volume} {68}},\
	\bibinfo {pages} {2849} (\bibinfo {year} {1990})}\BibitemShut {NoStop}%
\bibitem [{\citenamefont {Tan}\ \emph {et~al.}(1990)\citenamefont {Tan},
	\citenamefont {Snider}, \citenamefont {Chang},\ and\ \citenamefont
	{Hu}}]{Tan1990}%
\BibitemOpen
\bibfield  {author} {\bibinfo {author} {\bibfnamefont {I.-H.}\ \bibnamefont
		{Tan}}, \bibinfo {author} {\bibfnamefont {G.~L.}\ \bibnamefont {Snider}},
	\bibinfo {author} {\bibfnamefont {L.~D.}\ \bibnamefont {Chang}}, \ and\
	\bibinfo {author} {\bibfnamefont {E.~L.}\ \bibnamefont {Hu}},\ }\href
{\doibase http://dx.doi.org/10.1063/1.346245} {\bibfield  {journal} {\bibinfo
		{journal} {Journal of Applied Physics}\ }\textbf {\bibinfo {volume} {68}},\
	\bibinfo {pages} {4071} (\bibinfo {year} {1990})}\BibitemShut {NoStop}%
\bibitem [{\citenamefont {Li}\ \emph {et~al.}(2003)\citenamefont {Li},
	\citenamefont {Proskuryakov}, \citenamefont {Savchenko}, \citenamefont
	{Linfield},\ and\ \citenamefont {Ritchie}}]{Li2003}%
\BibitemOpen
\bibfield  {author} {\bibinfo {author} {\bibfnamefont {L.}~\bibnamefont
		{Li}}, \bibinfo {author} {\bibfnamefont {Y.~Y.}\ \bibnamefont
		{Proskuryakov}}, \bibinfo {author} {\bibfnamefont {A.~K.}\ \bibnamefont
		{Savchenko}}, \bibinfo {author} {\bibfnamefont {E.~H.}\ \bibnamefont
		{Linfield}}, \ and\ \bibinfo {author} {\bibfnamefont {D.~A.}\ \bibnamefont
		{Ritchie}},\ }\href {\doibase 10.1103/PhysRevLett.90.076802} {\bibfield
	{journal} {\bibinfo  {journal} {Phys. Rev. Lett.}\ }\textbf {\bibinfo
		{volume} {90}},\ \bibinfo {pages} {076802} (\bibinfo {year}
	{2003})}\BibitemShut {NoStop}%
\bibitem [{\citenamefont {Dmitriev}\ \emph {et~al.}(2012)\citenamefont
	{Dmitriev}, \citenamefont {Mirlin}, \citenamefont {Polyakov},\ and\
	\citenamefont {Zudov}}]{Dmitriev2012}%
\BibitemOpen
\bibfield  {author} {\bibinfo {author} {\bibfnamefont {I.~A.}\ \bibnamefont
		{Dmitriev}}, \bibinfo {author} {\bibfnamefont {A.~D.}\ \bibnamefont
		{Mirlin}}, \bibinfo {author} {\bibfnamefont {D.~G.}\ \bibnamefont
		{Polyakov}}, \ and\ \bibinfo {author} {\bibfnamefont {M.~A.}\ \bibnamefont
		{Zudov}},\ }\href {\doibase 10.1103/RevModPhys.84.1709} {\bibfield  {journal}
	{\bibinfo  {journal} {Rev. Mod. Phys.}\ }\textbf {\bibinfo {volume} {84}},\
	\bibinfo {pages} {1709} (\bibinfo {year} {2012})}\BibitemShut {NoStop}%
\bibitem [{\citenamefont {Stormer}\ \emph {et~al.}(1990)\citenamefont
	{Stormer}, \citenamefont {Pfeiffer}, \citenamefont {Baldwin},\ and\
	\citenamefont {West}}]{Stormer1990}%
\BibitemOpen
\bibfield  {author} {\bibinfo {author} {\bibfnamefont {H.~L.}\ \bibnamefont
		{Stormer}}, \bibinfo {author} {\bibfnamefont {L.~N.}\ \bibnamefont
		{Pfeiffer}}, \bibinfo {author} {\bibfnamefont {K.~W.}\ \bibnamefont
		{Baldwin}}, \ and\ \bibinfo {author} {\bibfnamefont {K.~W.}\ \bibnamefont
		{West}},\ }\href {\doibase 10.1103/PhysRevB.41.1278} {\bibfield  {journal}
	{\bibinfo  {journal} {Phys. Rev. B}\ }\textbf {\bibinfo {volume} {41}},\
	\bibinfo {pages} {1278} (\bibinfo {year} {1990})}\BibitemShut {NoStop}%
\bibitem [{\citenamefont {Mendez}\ \emph {et~al.}(1984)\citenamefont {Mendez},
	\citenamefont {Price},\ and\ \citenamefont {Heiblum}}]{Mendez1984}%
\BibitemOpen
\bibfield  {author} {\bibinfo {author} {\bibfnamefont {E.~E.}\ \bibnamefont
		{Mendez}}, \bibinfo {author} {\bibfnamefont {P.~J.}\ \bibnamefont {Price}}, \
	and\ \bibinfo {author} {\bibfnamefont {M.}~\bibnamefont {Heiblum}},\ }\href
{\doibase http://dx.doi.org/10.1063/1.95178} {\bibfield  {journal} {\bibinfo
		{journal} {Applied Physics Letters}\ }\textbf {\bibinfo {volume} {45}},\
	\bibinfo {pages} {294} (\bibinfo {year} {1984})}\BibitemShut {NoStop}%
\bibitem [{\citenamefont {Hirakawa}\ and\ \citenamefont
	{Sakaki}(1986)}]{Hirakawa1986}%
\BibitemOpen
\bibfield  {author} {\bibinfo {author} {\bibfnamefont {K.}~\bibnamefont
		{Hirakawa}}\ and\ \bibinfo {author} {\bibfnamefont {H.}~\bibnamefont
		{Sakaki}},\ }\href {\doibase 10.1103/PhysRevB.33.8291} {\bibfield  {journal}
	{\bibinfo  {journal} {Phys. Rev. B}\ }\textbf {\bibinfo {volume} {33}},\
	\bibinfo {pages} {8291} (\bibinfo {year} {1986})}\BibitemShut {NoStop}%
\bibitem [{\citenamefont {Kawamura}\ and\ \citenamefont
	{Das~Sarma}(1992)}]{Kawamura1992}%
\BibitemOpen
\bibfield  {author} {\bibinfo {author} {\bibfnamefont {T.}~\bibnamefont
		{Kawamura}}\ and\ \bibinfo {author} {\bibfnamefont {S.}~\bibnamefont
		{Das~Sarma}},\ }\href {\doibase 10.1103/PhysRevB.45.3612} {\bibfield
	{journal} {\bibinfo  {journal} {Phys. Rev. B}\ }\textbf {\bibinfo {volume}
		{45}},\ \bibinfo {pages} {3612} (\bibinfo {year} {1992})}\BibitemShut
{NoStop}%
\bibitem [{\citenamefont {Stern}(1972)}]{Stern1972}%
\BibitemOpen
\bibfield  {author} {\bibinfo {author} {\bibfnamefont {F.}~\bibnamefont
		{Stern}},\ }\href {\doibase 10.1103/PhysRevB.5.4891} {\bibfield  {journal}
	{\bibinfo  {journal} {Phys. Rev. B}\ }\textbf {\bibinfo {volume} {5}},\
	\bibinfo {pages} {4891} (\bibinfo {year} {1972})}\BibitemShut {NoStop}%
\bibitem [{\citenamefont {Serre}\ \emph {et~al.}(1989)\citenamefont {Serre},
	\citenamefont {Ghazali},\ and\ \citenamefont {Gold}}]{Serre1989}%
\BibitemOpen
\bibfield  {author} {\bibinfo {author} {\bibfnamefont {J.}~\bibnamefont
		{Serre}}, \bibinfo {author} {\bibfnamefont {A.}~\bibnamefont {Ghazali}}, \
	and\ \bibinfo {author} {\bibfnamefont {A.}~\bibnamefont {Gold}},\ }\href
{\doibase 10.1103/PhysRevB.39.8499} {\bibfield  {journal} {\bibinfo
		{journal} {Phys. Rev. B}\ }\textbf {\bibinfo {volume} {39}},\ \bibinfo
	{pages} {8499} (\bibinfo {year} {1989})}\BibitemShut {NoStop}%
\bibitem [{\citenamefont {Altshuler}\ and\ \citenamefont
	{Aronov}(1985)}]{Altshuler1985}%
\BibitemOpen
\bibfield  {author} {\bibinfo {author} {\bibfnamefont {B.}~\bibnamefont
		{Altshuler}}\ and\ \bibinfo {author} {\bibfnamefont {A.}~\bibnamefont
		{Aronov}},\ }in\ \href {\doibase
	http://dx.doi.org/10.1016/B978-0-444-86916-6.50007-7} {\emph {\bibinfo
		{booktitle} {Electron-Electron Interactions in Disordered Systems}}},\
\bibinfo {series} {Modern Problems in Condensed Matter Sciences},
Vol.~\bibinfo {volume} {10},\ \bibinfo {editor} {edited by\ \bibinfo {editor}
	{\bibfnamefont {A.}~\bibnamefont {Efros}}\ and\ \bibinfo {editor}
	{\bibfnamefont {M.}~\bibnamefont {Pollak}}}\ (\bibinfo  {publisher}
{Elsevier},\ \bibinfo {year} {1985})\ pp.\ \bibinfo {pages} {1 --
	153}\BibitemShut {NoStop}%
\bibitem [{\citenamefont {Lee}\ and\ \citenamefont
	{Ramakrishnan}(1985)}]{Lee1985}%
\BibitemOpen
\bibfield  {author} {\bibinfo {author} {\bibfnamefont {P.~A.}\ \bibnamefont
		{Lee}}\ and\ \bibinfo {author} {\bibfnamefont {T.~V.}\ \bibnamefont
		{Ramakrishnan}},\ }\href {\doibase 10.1103/RevModPhys.57.287} {\bibfield
	{journal} {\bibinfo  {journal} {Rev. Mod. Phys.}\ }\textbf {\bibinfo {volume}
		{57}},\ \bibinfo {pages} {287} (\bibinfo {year} {1985})}\BibitemShut
{NoStop}%
\bibitem [{\citenamefont {Zala}\ \emph
	{et~al.}(2001{\natexlab{a}})\citenamefont {Zala}, \citenamefont {Narozhny},\
	and\ \citenamefont {Aleiner}}]{Zala2001}%
\BibitemOpen
\bibfield  {author} {\bibinfo {author} {\bibfnamefont {G.}~\bibnamefont
		{Zala}}, \bibinfo {author} {\bibfnamefont {B.~N.}\ \bibnamefont {Narozhny}},
	\ and\ \bibinfo {author} {\bibfnamefont {I.~L.}\ \bibnamefont {Aleiner}},\
}\href {\doibase 10.1103/PhysRevB.64.214204} {\bibfield  {journal} {\bibinfo
	{journal} {Phys. Rev. B}\ }\textbf {\bibinfo {volume} {64}},\ \bibinfo
{pages} {214204} (\bibinfo {year} {2001}{\natexlab{a}})}\BibitemShut
{NoStop}%
\bibitem [{\citenamefont {Zala}\ \emph
	{et~al.}(2001{\natexlab{b}})\citenamefont {Zala}, \citenamefont {Narozhny},\
	and\ \citenamefont {Aleiner}}]{Zala2001a}%
\BibitemOpen
\bibfield  {author} {\bibinfo {author} {\bibfnamefont {G.}~\bibnamefont
		{Zala}}, \bibinfo {author} {\bibfnamefont {B.~N.}\ \bibnamefont {Narozhny}},
	\ and\ \bibinfo {author} {\bibfnamefont {I.~L.}\ \bibnamefont {Aleiner}},\
}\href {\doibase 10.1103/PhysRevB.64.201201} {\bibfield  {journal} {\bibinfo
	{journal} {Phys. Rev. B}\ }\textbf {\bibinfo {volume} {64}},\ \bibinfo
{pages} {201201} (\bibinfo {year} {2001}{\natexlab{b}})}\BibitemShut
{NoStop}%
\bibitem [{\citenamefont {Das~Sarma}\ and\ \citenamefont
	{Hwang}(2005)}]{DasSarma2005}%
\BibitemOpen
\bibfield  {author} {\bibinfo {author} {\bibfnamefont {S.}~\bibnamefont
		{Das~Sarma}}\ and\ \bibinfo {author} {\bibfnamefont {E.~H.}\ \bibnamefont
		{Hwang}},\ }\href {\doibase 10.1103/PhysRevLett.95.016401} {\bibfield
	{journal} {\bibinfo  {journal} {Phys. Rev. Lett.}\ }\textbf {\bibinfo
		{volume} {95}},\ \bibinfo {pages} {016401} (\bibinfo {year}
	{2005})}\BibitemShut {NoStop}%
\end{thebibliography}
\end{document}